\documentclass[pra,aps,showpacs,twocolumn,floatfix]{revtex4}
\usepackage{epsfig} \usepackage{graphics} \usepackage{bm}
\usepackage{amssymb}
\begin{document}
\title{Unconventional superfluidity of fermions in Bose-Fermi mixtures}
\author{O. Dutta$^1$ and M. Lewenstein$^{1,2}$}
\affiliation{${}^1$ ICFO-Institut de Ci{\`e}ncies Fot{\`o}niques, Mediterranean Technology Park, 08860 Castelldefels (Barcelona), Spain,}
\affiliation{${}^2$ ICREA-Instituci{\`o} Catalana de Recerca i Estudis Avan\c{c}ats, Lluis Companys 23, 08010 Barcelona, Spain.}

\date{\today}

\begin{abstract}
We examine two dimensional mixture of single-component fermions and
dipolar bosons. We calculate the self-enregies of the fermions in the normal state and the Cooper pair channel by including first order vertex correction to derive a modified Eliashberg equation. We predict appearance of superfluids with various non-standard pairing symmetries at experimentally feasible
transition temperatures within the strong-coupling limit of the Eliashberg equation. Excitations in these superfluids are
anyonic and follow non-Abelian statistics.
\end{abstract}

\maketitle
\section{Introduction}
Experimental realization of fermionic superfluidity in a quantum
degenerate ultracold gas \cite{sup} started a renewed interest in
the field. Attraction between fermionic particles favours pairing of
fermions resulting in superfluidity of the system. The paired
fermions, known as Cooper pairs, can have different kinds of internal
symmetries. The common ones found in nature have $s$-wave and
$d$-wave internal structure and conserve parity and time reversal
symmetry. In superfluid $^{3}$He, the so-called $A$ and $A_1$ phases
are characterized by Cooper pairs with nonzero magnetic
orbital momenta. Also, there was a theoretical prediction of higher
order $f$-wave pairing in $^{3}$He \cite{bam, mer}. Cooper pairs
with chiral $p_x+i p_y$-wave internal structure are believed to be
responsible for the observed superfluidity of electrons in Strontium Ruthenate
\cite{str}. This kind of pairing breaks the time-reversal symmetry.
Spinless chiral $p$-wave superfluid state has formal resemblance to
the ``Pfaffian" state proposed in relation to the fractional quantum
Hall state with filling factor $5/2$ \cite{QH1, CN1, read1}. When
confined in a two dimensional geometry, excitations in the chiral
$p$-wave superfluid become so called non-Abelian anyons. Anyons are
particles living in a two dimensional plane that under exchange do
not behave neither as bosons nor as fermions. For non-Abelian
anyons, exchange of two such particles depends on the order of the
exchange \cite{W1, W2, W3}. Apart from fundamental interest in the
existence of such particles, non-Abelian anyons find remarkable
applications in the field of quantum information for quantum
memories and fault-tolerant quantum computation \cite{K1}.

               Recently it has been shown that quasi-particles in
vortex excitations of chiral two-dimensional $p$-wave spinless
superfluids obey non-Abelian statistics \cite{iva, CN2}. Using
$p$-wave Feshbach resonances in fermionic ultracold atoms, such
superfluids can be realized in principle, but this procedure is very
difficult because of non-elastic loss processes \cite{gur1}. Another
proposal for $p$-wave superfluid with dipolar fermions has been
recently formulated in Ref \cite{coop} with transition temperature in
the order of $.1\epsilon_f$, where $\epsilon_f$ is the Fermi
energy. Bose-Fermi mixtures are another candidate for creating
superfluidity in fermions via boson mediated interactions
\cite{viv1} and have formal resemblance to phonon mediated
superconductivity in metals. It was found, however, when the bosons
and single-component fermions are completely mixed, increasing the boson-fermion
interaction strength, or fermionic density may induce dynamical instability of the condensate resulting 
in phase separation between the mixture \cite{viv1, pet, yang, taka}. Near phase
separation, inclusion of the dressing of phonons is predicted 
to increase the transition temperature. In
Ref.~\cite{bul}, the authors found that close to Feshbach
resonances, a Bose condensate of dimers can induce a strong pairing
in the $p$-wave channel.

         In the present paper, we discuss another way to generate high temperature superfluids
in a Bose-Fermi mixture. We study the property of superfluidity in
Bose-Fermi mixtures, where bosons are interacting via long-ranged
dipolar interactions. We show that the transition temperature for
$p$-wave superfluidity can become comparable to the Fermi energy.
More importantly, we find that other more exotic Cooper pairs with
$f$- and $h$-wave internal symmetries are possible in certain range
of Fermi energies without bosons and fermions separating. In
addition, we study the excitations in chiral states of the odd-wave
superfluids and point out their non-Abelian anyonic nature.
Experimentally, an available bosonic species, where prominent
dipolar interaction can be achieved using Feshbach resonance is
Cr$^{52}$ \cite{Pf1, Pf2}. Another route towards achieving ultracold
dipolar gas is to experimentally realize quantum degenerate
heteronuclear molecules \cite{Jin1}, which have permanent electric
moment. Thus, in the near future a quantum degenerate mixture of
dipolar bosons and fermions will be achievable experimentally.

In Section~\ref{dbec} we review the properties of a dipolar Bose
condensate in a pancake trap. We discuss specially the excitation
spectrum of such condensates. Then in Section~\ref{bfex} we study
the boson-fermion interaction and the many-body effect of fermions
on dressing the excitation spectrum of the condensate. In
Section~\ref{infer}, we discuss the interaction between the fermions
mediated by bosons in different angular momentum channel. 
By integrating out the bosonic mode, we show that the absolute value of the interaction between 
fermions in the $p$, $h$, $f$-wave angular momentum channel
are comparable depending on the Fermi energy. In Section~\ref{verc}, we go beyond
Migdal's limit and include first order vertex corrections to study
fermion self-energy in normal state as well as the mass renormalization function
in the high temperature limit. In doing so, we include the full effect of retardation and strong
momentum dependance of the bosonic excitation spectrum and bosonic
propagator. We find that vertex correction reduces the mass renormalization function. 
In Section~\ref{sverc} we calculate the self-energy in
Cooper-pair channel for $p$, $f$, $h$-wave order parameters
including the vertex correction and cross-interaction for
temperature $T>T_c$. By deriving a vertex-corrected strong-coupling
Eliashberg equation, we solve for transition temperatures in
different angular momentum channels. In Section~\ref{abex}, we present a brief
discussion regarding the possible occurrence of non-Abelian Majorana
fermions for broken time-reversal $p$, $f$, and $h$-wave
superfluids. We solve the Bugoliubov-deGennes equation for the $p$,
$f$, and $h$-wave superfluids in the limit of large distance to find
the Majorana bound states.

 \section{Dipolar Bose-Einstein condensate}\label{dbec}
       Our system consists of dipolar bosons mixed with single component
fermions, confined in a quasi-two dimensional geometry by a harmonic
potential in the $z$ direction with the condition
$m_b\omega^2_b=m_f\omega^2_f=$ where $m_b$ and $m_f$ is the mass of
bosons and fermions respectively and $\omega_b, \omega_f$ is the
trapping frequency. First, we assume that the bosons are polarized
along the $z$ direction. The dipolar interaction reads $ V_{\rm
dd}=\frac{4\pi g_{\rm dd}}{3}( 3k^2_z/k^2-1 )$ in momentum space,
where $g_{\rm dd}$ is the dipole-dipole interaction strength. For
atoms $g_{\rm dd}= \mu_0 \mu^2_m/4\pi $, and for dipolar molecules
$g_{\rm dd}=\mu^2_e/4\pi\epsilon_0 $, where $\mu_m$ and $\mu_e$ are
the magnetic moment of the atoms and the electric dipole moment of
the molecules, respectively. We assume that the $z$ dependence of
bosonic density is given by a Thomas-Fermi profile,
$$
n_b(x,y,z)=\frac{3n_b}{4R_z} \left (1-\frac{z^2}{R^2_z} \right ),
$$
where the Thomas-Fermi radius $R_z$ is determined variationally.
After integrating over $z$ dependence of the density profile of
bosons, the total interaction takes the form $ V_{\rm eff}=\frac{8
\pi g_{\rm dd}}{5R_z} \mathcal{V}(\tilde{k}_{\bot})$ where
\begin{eqnarray}
\mathcal{V}(\tilde{k}_{\bot})&=& \frac{3g}{8\pi g_{\rm
dd}}-\frac{1}{2} \nonumber\\
 &+&  \frac{15[ 2\tilde{k}^3_{\bot} - 3\tilde{k}^2_{\bot} -
3(1+\tilde{k}_{\bot})^2\exp(-2\tilde{k}_{\bot})+3]}{8
\tilde{k}^5_{\bot}}, \nonumber\\
\end{eqnarray}
$\tilde{k}_{\bot}=k_{\bot}R_z$ and $g$ is the contact interaction
between the bosons. $\mathcal{V}(k_{\bot})$ is repulsive for small momentum and can be
attractive in the high momentum limit depending on the contact
interaction. Subsequently, we write the Hamiltonian of the
dipolar bosons in the condensed phase, $H_b = \sum_{\vec{k_{\bot}}}
\Omega_0(\vec{k_{\bot}}) \beta^{\dagger}_{\vec{k_{\bot}}}
\beta_{\vec{k_{\bot}}}$, where $\beta^{\dagger}_{\vec{k_{\bot}}}$
and $\beta_{\vec{k_{\bot}}}$ are Bugoliubov operators. The
excitation frequency $\Omega_0(\vec{k_{\bot}})$ is given in the
units of trap frequency,
\begin{equation}\label{cex1}
\Omega^2_0({k}_{\bot}\ell_0)=\frac{[k_{\bot}\ell_0]^4}{4}+ g_{\rm
3d}\frac{\ell_0}{R_z} \mathcal{V}\left ( {k}_{\bot}\ell_0
\frac{R_z}{\ell_0} \right ) [k_{\bot}\ell_0]^2,
\end{equation}
where $\ell_0=\sqrt{\hbar/m_b \omega_b}$. We define a dimensionless
dipolar interaction strength $g_{\rm 3d}=8\pi m_b g_{\rm dd} n_b
\ell_0/5\hbar^2$ which will be used later, where $\ell_0$ is the
ground state oscillator length. Also the phonon propagator $D_0(i
\omega_s, k_{\bot})$ in this regime is given by
\begin{equation}
D_0(i \omega_s, k_{\bot})=-\frac{\Omega_0({k}_{\bot})}{\omega^2_s+\Omega^2_0({k}_{\bot})},
\end{equation}
where bosonic Matsubara frequency $\omega_s=2s$ where $s$ is an integer.
By minimizing the mean field energy of the Bose condensate within
Thomas-Fermi regime we find that
$$
R_z/\ell_0 = \left ( 2.5 g_{\rm 3d} \left [ 1 + \frac{3}{16\pi
g_{\rm dd}} \right ] \right )^{1/3}.
$$
For $\frac{3g}{4\pi g_{\rm dd}} > 1$, the excitation
spectrum of the condensate, denoted by $\Omega(\vec{k_{\bot}})$, can
be divided into two parts: i) $\sim k_{\bot}$, phonon spectrum for
small momenta and ii) $\sim k^2_{\bot}$, free-particle like spectrum
for higher momenta \cite{pit}. For $\frac{3g}{4\pi g_{\rm dd}}
<1$ and $g_{\rm 3d}$ greater than a critical value,
$\Omega(\vec{k_{\bot}})$ has a minimum at momentum $\tilde{k}_0$
\cite{Lew1}, where $\tilde{k}_0$ is in intermediate momentum regime.
Following Landau, the excitations around the minimum are called
``rotons''. With increasing $g_{\rm 3d}$ the excitation energy at
$\tilde{k}_0$ decreases and eventually vanishes for a critical
particle density. When the particle density exceeds that critical
value, the excitation energy becomes imaginary at finite momentum
and the condensate becomes unstable. The use of Thomas-Fermi density
profile is justified in this region as the chemical potential
necessary to reach roton instability exceeds $\hbar \omega_z$.

\section{Boson-fermion interaction and dressed excitation}\label{bfex}
   In this section, first we disccuss the condensate-fermion interaction Hamiltonian and dynamical stability
of the condensate. We find the the dressed propagator for the Bugoliubov quasi-particles
in presence of fermionic particle hole excitation within second-order perturbation
theory. We then discuss the appearance of roton instability in the dressed excitation spectrum of the condensate. 

     Kinetic energy for the single component non-interacting fermions, is
characterized by the Hamiltonian $H_f = \sum_{\vec{k_{\bot}}} \left
[ \xi(\vec{k_{\bot}}) - \epsilon_f \right ]
c^{\dagger}_{\vec{k_{\bot}}} c_{\vec{k_{\bot}}}$, where
$c^{\dagger}_{\vec{k_{\bot}}}$ and $c_{\vec{k_{\bot}}}$ are
fermionic creation and destruction operator. $\xi(\vec{k_{\bot}}) =
\vec{k}^2/2m_f- \epsilon_f$ is the dispersion energy of the fermions
and $\epsilon_f$ is the Fermi energy. The density profile of fermions
along the $z$ direction is approximated by a normalized Gaussian
with width $\ell_f=\sqrt{\hbar/m_f\omega_f}$. The fermions are
interacting with the bosons via short ranged contact interaction of
strength $g_{\rm bf}$. After integrating over the $z$ coordinate,
the boson-fermion interaction Hamiltonian reads
\begin{equation}
 H_{\rm bf}=\frac{3g_{\rm bf}\alpha}{4\sqrt{\pi}
R_z}\alpha\sum_{\vec{k_{\bot}}, \vec{q_{\bot}}}
c^{\dagger}_{\vec{k_{\bot}}}
c_{\vec{k'_{\bot}}}b^{\dagger}_{\vec{q'_{\bot}}}b_{\vec{q_{\bot}}},
\end{equation}
where $b_{\vec{q_{\bot}}}, b^{\dagger}_{\vec{q_{\bot}}}$ are
bosonic creation and annihilation operator, and
$\alpha=\frac{\ell_f}{R_z}\exp \left (-\frac{R^2_z}{\ell^2_f} \right
) - \frac{\sqrt{\pi}}{2} \left ( \frac{\ell^2_f}{R^2_z}-2 \right )
{\rm erf} \left ( \frac{R_z}{\ell_f} \right ), $ with ${\rm
erf}(..)$ being the error function. Next, we concentrate our
attention on the interaction between the fermions and the
Bugoliubov quasi-particles. The fluctuations in fermionic density
couple to the density fluctuations present in the Bose condensate.
Due to the momentum dependence of the bosonic excitations, the
condensate-fermion interaction becomes function of momentum.
Including the fluctuations in the Bose condensate in the $x-y$
plane, the condensate-fermion interaction Hamiltonian can be written
as
\begin{equation}\label{hbf}
 H_{\rm bf}=\frac{3g_{\rm
bf}}{4\sqrt{\pi} R_z}\alpha\sum_{\vec{k_{\bot}}, \vec{q_{\bot}}}
\gamma(\vec{k_{\bot}}) c^{\dagger}_{\vec{k_{\bot}}}
c_{\vec{q_{\bot}}-\vec{k_{\bot}}}\left [ b_{\vec{k_{\bot}}} +
b^{\dagger}_{-\vec{k_{\bot}}} \right ],
\end{equation}
where the momentum dependent coupling constant is given by
$\gamma(\vec{k_{\bot}})=\sqrt{2n_b\epsilon_b(\vec{k_{\bot}})/\Omega_0(\vec{k_{\bot}})}$.
%----------------------------------------
\begin{figure}[ht]
\begin{center} \epsfig{file=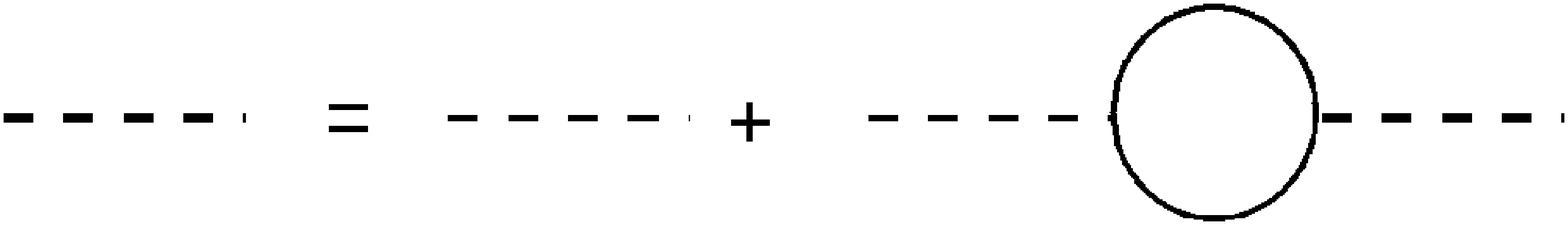,width=5.5cm, height=1.2cm} \caption{\label{bprop}
Schematic diagram of phonon propagator including the polarizing effect of the fermions. The thick dash-dotted line corresponds to the dressed phonon propagator $D(i \omega_s, k_{\bot})$, while the thin dash-dotted line corresponds to non-interacting phonon propagator $D_0(i \omega_s, k_{\bot})$. The solid line corresponds to free fermion propagator.}
\end{center}
\end{figure}
%------------------------------------------
Due to the interaction between Bugoliubov quasi-particles and
fermions, as shown in Fig.~\ref{bprop}, the bosonic excitations are
dressed by the fermions resulting in dressing of bare condensate
excitation spectrum in Eq.~(\ref{cex1}). This can be derived
within second order perturbation theory as shown in
Fig.~\ref{bprop}. Subsequently the new phonon propagator is given by
\begin{equation}\label{dph}
D(i \omega_s, k_{\bot})=-\frac{\Omega_0({k}_{\bot})}{\omega^2_s+\Omega^2({k}_{\bot})},
\end{equation}
where the dressed excitation spectrum $ \Omega({k}_{\bot}\ell_0)$
for the dressed Bugoliubov quasi-particles is expressed by
\begin{equation}\label{cex2}
\Omega(\vec{k_{\bot}}) = \Omega_0(\vec{k_{\bot}})
\sqrt{1-\frac{|\gamma(\vec{k_{\bot}})|^2N_0}{\Omega_0({k}_{\bot})} h(\omega, k_{\bot})},
\end{equation}
where the two dimensional Lindhard function
$$
h(\omega,
k_{\bot})=1-|\omega|\frac{\theta(|\omega|-v_fk_{\bot})}{\sqrt{\omega^2-v^2_fk^2_{\bot}}},
$$
and $\omega$ is the transfer of energy. As we are interested in
Cooper instability of the fermions which happens for momenta close
to Fermi momentum, the transfer of energy is of the order of $\omega
\ll \epsilon_f$. In this limit the Lindhard function $h(\omega,
k_{\bot})=1$. For future use, we define an effective interaction
strength between the bosons and fermions,
$$
\mathcal{G}_{\rm bf}=\frac{45g^2_{\rm bf}N_0}{128\pi g_{\rm dd}
\ell_0}\alpha^2.
$$

\begin{figure}[ht]
\begin{center} \epsfig{file=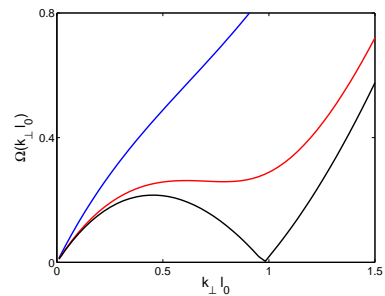,width=5cm} \caption{\label{excc}
The dressed excitation spectrum $\Omega(\vec{k_{\bot}})$ as a function of $\vec{k_{\bot}}$
For various values of the Bose-Fermi interaction $\mathcal{G}_{\rm bf}=0$(blue line), $.51$(red line), $.57$(black line).
The fixed parameters are $m_f=53$, $\frac{3g}{4\pi g_{\rm dd}}=.6$, and $g_{\rm 3d}=2.5$}
\end{center}
\end{figure}

   In order to determine dynamical stability of the uniform condensate, we study the properties of the
dressed spectrum of Bugoliubov quasi-particles in the parameter
regime of $\frac{3g}{4\pi g_{\rm dd}} > 1$, where the bare spectrum has
no roton minimum. From Eq.~(\ref{cex2}), due to the attractive effect
of Bose-Fermi interaction, we notice that roton minimum in the
excitation spectrum $\Omega(\vec{k_{\bot}})$ can develop for certain
critical interaction strength $g_{\rm 3d}$ and $G_{\rm bf}$.
Consequently, the uniform condensate will become instable as the
excitation spectrum becomes imaginary at a finite momentum. But, as
long as the roton gap remains positive, the uniform state of the
bosons will be stable or metastable \cite{od1}. In Fig.~\ref{excc},
we have plotted the dressed excitation spectrum for different values
of Bose-Fermi interaction $G_{\rm bf}$. With no boson-fermion
interaction, as $\frac{3g}{4\pi g_{\rm dd}} > 1$, the excitation
spectrum has the usual phonon regime for low momenta and free
particle regime for higher momenta. For a critical $G_{\rm bf}$,
i.e. $G_{\rm bf}\approx.5$ in Fig.~\ref{excc}, in the region of
%--------------------------------
%---------------------------------
intermediate momenta, the excitation spectrum have roton-maxon
character. With further increase of $G_{\rm bf}$, the roton gap goes
to zero and the excitation spectrum becomes imaginary at a finite
momentum. Subsequently the bosons undergo a phase transition to a
state with periodic dencity modulation, which finally collapses. This periodic density
wave state can be stabilized for repulsive contact interaction
obeying $3g/8\pi g_{\rm dd}\approx \mathcal{G}_{\rm bf}
\frac{\ell_0}{R_z}$ \cite{od1}. The appearance of roton minimum in dressed
excitation spectrum in this case is entirely due to the many-body
effect of the fermions on Bugoliubov quasi-particles. Another point we like to stress is that,
the roton instability is always reached before the phonon instability pointing towards phase separation \cite{viv1}.

\section{Effective interaction between fermions}\label{infer}
       In this section we find the interaction between the fermions mediated by the 
Bugoliubov quasi-particles in the limit of $T=0$. We then obtain the interaction strength in different angular momentum channel,
which varies as a function of $dimensionality$ parameter, defined as
$\eta=\epsilon_f/\hbar\omega_f$. 

   Integrating out the bosonic degree
of freedom in Eq.~(\ref{hbf}), and inclusion of the effect of dressed
excitation spectrum, results in effective interaction between the
fermions \cite{wang1},
\begin{equation}\label{vph}
V_{\rm ph}(\vec{q_{\bot}}, i\omega_s)= \frac{9g^2_{\rm
bf}\alpha^2}{16\pi R^2_z}
 \frac{n_b q^2_{\bot}/m_b}{\omega^2-\Omega^2(\vec{q_{\bot}})},
\end{equation}
where ${q}^2_{\bot}=2k^2_f(1-\cos\phi)$, is the momentum exchange
between the interacting particles along the Fermi surface. At $T=0$, assuming momentum
transfer occur around Fermi momentum $k_f$, we can
expand Eq.~(\ref{vph}) as $ V_{\rm ph}(\vec{k_{\bot}}, 0)=
- \sum_{L=...,-1,0,1,...}\lambda_L(0,0) e^{iL\phi}$,
and the dimensionless effective interaction between the fermions in
angular momentum channel $L$ is given by
\begin{eqnarray}\label{lnv1}
\lambda_L(0,0) &=& \mathcal{G}_{\rm bf} \int^{2\pi}_0 \frac{\exp [ i L \phi ]
d\phi/2\pi}
{\frac{\eta R^2_z}{g_{\rm 3d} \ell^2_f}(1-\cos \phi)+\frac{R_z}{\ell_0}\mathcal{V}(\sqrt{\frac{R_z}{{\ell_f}}}\eta(1-\cos\phi))}. \nonumber\\
\end{eqnarray}
As we are considering single component fermions, pairing will occur
in the odd angular momentum channels.
%--------------------------------
\begin{figure}[ht]
\begin{center} \epsfig{file=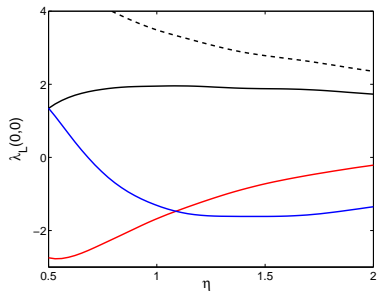,width=5cm} \caption{\label{eta53l}
Figure of effective fermion-fermion interaction strength $\lambda_L(0,0)$ in the angular momentum channels $L=0$(dashed line), $L=1$(black
line),$=3$(red line),$=5$(blue line) as a function of the fermion
dimensionality parameter $\eta$. We fixed $mf/mb=53/52$, $g_{\rm 3d}=4, 3g/8\pi g_{\rm dd}=0.6$,
and $\mathcal{G}_{\rm bf}=0.5$.}
\end{center}
\end{figure}
%---------------------------------

         Next, we look into the variation of $\lambda_L(0, 0)$ as a function of
$\eta$. For concreteness, we assume a $^{52}$Cr-$^{53}$Cr mixture.
The interaction strengths from Eq.~(\ref{lnv1}) in various angular
momentum channels have been plotted in Fig.~\ref{eta53l}. We find that
with changing dimensionality, the strengths in different angular
momentum channel varies. Additionally, $|\lambda_1(0, 0)| \sim
|\lambda_3(0, 0)| ~\sim |\lambda_5(0, 0)|$, and depending on the
dimensionality they can be positive or negative.

\section{Fermionic self-energy including vertex correction}\label{verc}
    In this section, we consider the fermion self-energy in the normal
state due to the interaction between the fermions and dressed
Bugoliubov quasi-particles. In doing so, we explicitly take into
account the momentum dependance as well as the retardation of the
phonon propagator. More importantly, we also go beyond Migdal's
adiabatic limit and take into account the effect of vertex
correction. This kind of non-adiabatic correction is introduced to
electron-phonon system in metals in Refs \cite{gps, gps1, gps2} in
the limit of $T \rightarrow 0$.  Our main result is that as temperature gets lower, the vertex-corrected mass renormalization function $Z(T)$ also gets smaller.

\begin{figure}[ht]
\begin{center} \epsfig{file=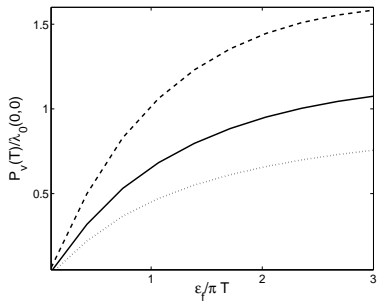,width=5cm, height=4cm} \caption{\label{pvf}
Plot of the vertex correction in $P_v$ as a function of $\frac{\epsilon_f}{\pi T}$ for different
values of the dimensionality parameter $\eta=1.5$(the dotted line),
$\eta=1$(the solid line), $\eta=.6$(the dashed line). We fixed
$m_f/m_b=53/52$, $\frac{3g}{4\pi g_{\rm
dd}}=.6,\mathcal{G}_{\bf}=.5$, and $ g_{\rm 3d}=3.8$.}
\end{center}
\end{figure}

     The unperturbed Green's function for
the fermions is given by $G_0(\vec{k_{\bot}}, i
\omega_n)=1/(i\omega_n - \xi(\vec{k_{\bot}}))$, where the Matsubara
frequency $\omega_n=(2n+1)\pi T$, $n$ being integer.
%--------------------------------
\begin{figure}[ht]
\begin{center} \epsfig{file=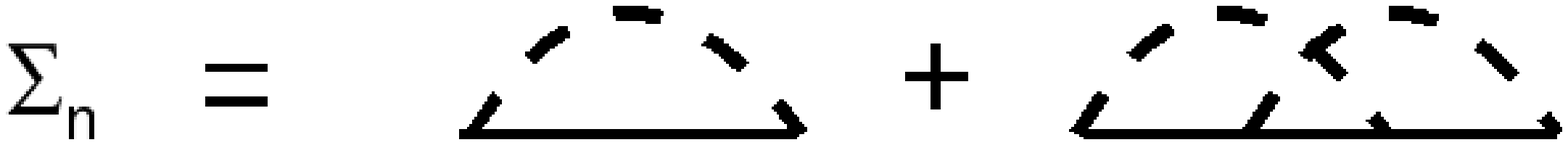,width=6cm} \caption{\label{ver}
Figure of fermion self-energy in normal state including the first order vertex correction.
The solid line denotes the fermion propagator while the thick dashed line denotes the dressed phonon propagator of Eq.~(\ref{dph})}
\end{center}
\end{figure}
%---------------------------------
The normal self-energy $\Sigma_n(\vec{k_{\bot}})$, including the
first order vertex correction, as shown in Fig.~\ref{ver}, is given
by
\begin{eqnarray}\label{eq1}
\Sigma_n(\vec{k_{\bot}}, i\omega_n) &=& \Sigma^1_n(\vec{k_{\bot}},
i\omega_n) + \Sigma^v_n(\vec{k_{\bot}}, i\omega_n),
\end{eqnarray}
where $\Sigma^1_n(\vec{k_{\bot}}, i\omega_n)$ comes from the first diagram in Fig.~\ref{ver} and $\Sigma^v_n(\vec{k_{\bot}}, i\omega_n)$ arises from the second diagram which denotes the vertex correction. Also,
\begin{widetext}
\begin{eqnarray}\label{nv1}
\Sigma^1_n(\vec{k_{\bot}}, i\omega_n) &=& -T\sum_{m,\vec{q_{\bot}}}
|\gamma(\vec{k_{\bot}}-\vec{q_{\bot}})|^2 D(\omega_m-\omega_n,
\vec{k_{\bot}}-\vec{q_{\bot}}) G_0(i\omega_m,
\vec{q_{\bot}}), \nonumber\\
 \label{nv2}\Sigma^v_n(\vec{k_{\bot}}, i\omega_n)
&=& T^2\sum_{m,l, \vec{q_{\bot}},\vec{p_{\bot}}}
|\gamma(\vec{k_{\bot}}-\vec{q_{\bot}})|^2
 |\gamma(\vec{k_{\bot}}-\vec{p_{\bot}})|^2 D(\omega_l-\omega_n,
\vec{k_{\bot}}-\vec{p_{\bot}})D(\omega_m-\omega_n,
\vec{k_{\bot}}-\vec{q_{\bot}})  \\
&\times& G_0(i\omega_m, \vec{q_{\bot}}) G_0(i\omega_l, \vec{p_{\bot}})
G_0(i(\omega_l-\omega_n+\omega_m),
\vec{p_{\bot}}-\vec{k_{\bot}}+\vec{q_{\bot}}), \nonumber\\
\end{eqnarray}
\end{widetext}
By taking the average over Fermi energy($|\vec{k_{\bot}}| \approx
|\vec{q_{\bot}}| \approx k_f$), and denoting $\xi(\vec{k_{\bot}})=
\hbar^2 k^2_{\bot}/2m - \epsilon_f = x$, $\xi(\vec{p_{\bot}})=
\hbar^2 k^2_{\bot}/2m - \epsilon_f = x'$, and
$\xi(\vec{p_{\bot}}-\vec{k_{\bot}} \vec{q_{\bot}}) -\mu = x' +
2\epsilon_f \alpha$, where
\begin{equation}
\alpha = 1-\cos(\phi')+\cos(\theta)-\cos(\gamma)
\end{equation}
and the angles $(\vec{k_{\bot}},\vec{q_{\bot}})=\phi'$,
$(\vec{k_{\bot}},\vec{p_{\bot}})=\theta$, $\gamma=\phi'-\theta$. Using Euler-Mclauren
summation formula, we can transform the sum over $l$ in Eq.~(\ref{nv2}) to
integral as,
\begin{widetext}
\begin{eqnarray}
P(x',m,T,\phi',\theta) &=& -\sum_{l} D(\omega_l-\omega_n,
\vec{k_{\bot}}-\vec{p_{\bot}}) \\
&\times& G(i\omega_m, \vec{q_{\bot}}) G(i\omega_l, \vec{p_{\bot}})
G(i(\omega_l-\omega_n+\omega_m),
\vec{p_{\bot}}-\vec{k_{\bot}}+\vec{q_{\bot}}) \nonumber\\
&=& \int^{+\infty}_{-\infty} \frac{d\omega}{2\pi}
\frac{\Omega^2}{\omega^2+\Omega^2} \frac{1}{i\omega - x'+i\pi T}
\frac{1}{(\omega+2m\pi T) - x'-\alpha+i\pi T}
\nonumber\\
&=& \frac{i\Omega}{2(2m\pi T - i\alpha)} \left
[\frac{\theta(x')}{\Omega+x'-i\pi T} -
\frac{\theta(-x')}{\Omega-x'+i\pi T} \right . \nonumber\\
&-& \left . \frac{\theta(x'+\alpha)}{\Omega-2i m\pi T+x'-i\pi T+\alpha}
+\frac{\theta(-x'-\alpha)}{\Omega+2i m\pi T-x'+i\pi T-\alpha} \right
], \nonumber\\
\end{eqnarray}
\end{widetext}
here $\theta(..)$ is the Step function. We can rewrite $\Sigma_n(i\omega_n)$ as
\begin {equation}
\Sigma_n(i\omega_n)=\chi(i\omega_n)+i\omega_n [1-Z(i\omega_n)],
\end {equation}
where $\chi(i\omega_n)$ is the real part of $\Sigma_n(i\omega_n)$ and $Z(i\omega_n)$ is known as the mass renormalization function.
$\chi(\pi T)$ shifts the bare fermion dispersion energy $\xi(\vec{k_{\bot}})$. As we are
considering strong-coupling superfluidity, we only consider the term
corresponds to $n=0$ \cite{mahan}. Then the mass renormalization
function $Z(\pi T)$ and $\chi(\pi T)$ in the conventional Eliashberg form is given by,
\begin{eqnarray}\label{z11}
Z(\pi T) &=& 1 + \lambda_Z(T), \\
\chi(\pi T)&=&\sum_m {\rm sign}(\omega_m) {\rm Im}[P_v(m, T)],
\end{eqnarray}
 with an effective coupling constant
\begin{equation}\label{z1}
\lambda_Z(T)= \lambda_0(0, 0)-P_v(T),
\end{equation}
and $P_v(T)=\sum_m {\rm sign}(\omega_m) {\rm Re}[P_v(m, T)]$, where
\begin{widetext}
\begin{eqnarray}\label{pvm}
\lambda_L(s,T) &=& - \int^{2\pi}_0 |\gamma(\phi)|^2  D(\omega_s, \phi) \cos L\phi
d\phi/ 2\pi, \nonumber\\
P_v(m, T)&=& \int^{\epsilon_f}_{-\epsilon_f} dx'\int^{\pi}_{-\pi}
\frac{d\phi'}{2\pi} d\theta |\gamma(\phi')|^2
 |\gamma(\theta)|^2 P(x',m,T,\phi',\theta), \nonumber\\
\end{eqnarray}
\end{widetext}
and ${\rm Re}[f]$, ${\rm Im}[f]$ denotes the real and imaginary part of the function $f$. 
       In Fig.~\ref{pvf} we plot $P_v(T)$ as a function of temperature
for various values of dimensionality $\eta$. From the expression of $\lambda_L(s, T)$ in Eq.~(\ref{pvm}), we notice that
$\lambda_L(0, 0)$ is always positive as shown in Fig.~\ref{eta53l}. $P_v(T)/\lambda_0(0, 0)$ is the expansion parameter
for the perturbative scheme. The correction
$P_v(T)$ is always found to be positive, which reduces the
coupling strength $\lambda_Z(T)$. With decreasing temperature,
$P_v(T)/\lambda_0(0, 0)$ increases before saturating. This saturation
value increases with decreasing dimensionality as seen in
Fig.~\ref{pvf}. For smaller $\eta$, $P_v(T)\lambda_0(0,0)$ becomes
higher than one for lower temperature, invalidating any perturbative
calculation in that low temperature region.
%--------------------------------
%---------------------------------

\begin{figure}[ht]
\begin{center} \epsfig{file=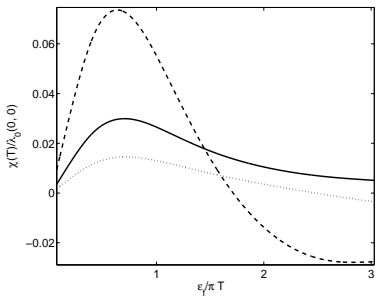,width=5cm, height=4cm} \caption{\label{chii}
Figure of $\chi(T)$ as a function of $\frac{\epsilon_f}{\pi T}$ for different
values of the dimensionality parameter $\eta=1.5$(the dotted line),
$\eta=1$(the solid line), $\eta=.6$(the dashed line). We fixed
$m_f/m_b=53/52$, $\frac{3g}{4\pi g_{\rm
dd}}=.6,\mathcal{G}_{\bf}=.5$, and $ g_{\rm 3d}=3.8$.}
\end{center}
\end{figure}

        Next, we calculate the real part of the fermion self energy $\chi(\pi T)$. Figure \ref{chii}
%--------------------------------
%---------------------------------
shows a such generic case for the same parameters as Fig.~\ref{pvf}. We find that $\chi(T) \ll \lambda_0(0,0)$ even in the 
temperature region with high $P_v(T)$. Henceforth, we can
neglect the effect of energy shift of the fermions due to the smallness of $\chi(T)$.

\section{Self-energy in Cooper-pair channel and Transition temperature}\label{sverc}

\begin{figure}[ht]
\begin{center} \epsfig{file=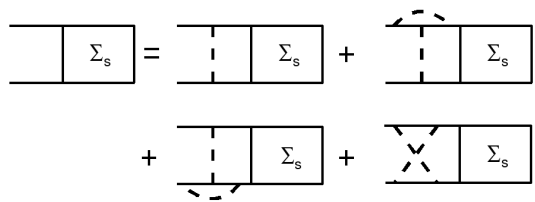, width=7cm} \caption{\label{sver}
Schematic diagram of fermion self-energy in the Cooper-pair channel
channel including the first order vertex correction. The solid line
denotes the fermion propagator while the thick dashed line denotes
the dressed phonon propagator of Eq.~(\ref{dph}). The first diagram
denotes the direct interaction, while the second and third diagram
denote the first order vertex corrections, and the third diagram
denotes the cross-term. }
\end{center}
\end{figure}

    In this section we look at the fermionic self-energy in
the Cooper-pair channel, $\Sigma_s(\vec{k_{\bot}}, i\omega_n)$ as represented
diagramatically in Fig.~\ref{sver}, close to transition temperature $T_c$. Then we find transition temperature within strong-coupling limit considering the terms $n=0, -1$ \cite{umm}. The
main results of this section are : $i)$ Depending on the angular momentum channel, 
the vertex-corrected interaction strength can increase as a function of dimensionality and temperature.
$ii)$ The solution of modified Eliashberg equation supports $T_c\sim .1\epsilon_f$ in the strong-coupling limit for
$p-$,$f-$, and $h-$-wave order parameters.

%--------------------------------
\begin{figure}[ht]
\begin{center} \epsfig{file=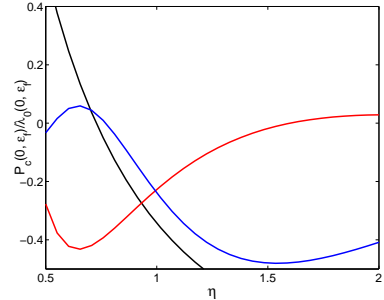,width=5cm} \caption{\label{pc53}
Plot of cross-term interaction $P_c(0,\epsilon_f)$ as a function of $\eta$ for $m_f/m_b=53/52$, $\pi T=\mu$, $g_{\rm 3d}=4.0$, $\mathcal{G}_{\rm bf}=0.5$ and $\frac{3g}{4\pi g_{\rm
dd}}=0.6$. The black, red and blue lines correspond to $p$, $f$, and $h$-wave angular momentum channel respectively. }
\end{center}
\end{figure}

%---------------------------------
For $T>T_c$, the fermion self-energy in the Cooper-pair channel,
$\Sigma_s(\vec{k_{\bot}})$, is given by
\begin{equation}
\label{s1}\Sigma_s(\vec{k_{\bot}}, i\omega_n)=
\Sigma^1_s(\vec{k_{\bot}}, i\omega_n) + \Sigma^v_s(\vec{k_{\bot}},
i\omega_n)+\Sigma^{c}_c(\vec{k_{\bot}}, i\omega_n),
\end{equation}
where $\Sigma^1_s(\vec{k_{\bot}}, i\omega_n)$ comes from the first diagram in Fig.~\ref{sver}, the second and third diagram gives equal contribution and denoted by $\Sigma^v_s(\vec{k_{\bot}}, i\omega_n)$
while the last diagram denoting cross interaction is given by $\Sigma^{c}_c(\vec{k_{\bot}}, i\omega_n)$, and
\begin{widetext}
\begin{eqnarray}\label{ss}
\Sigma^1_s(\vec{k_{\bot}}, i\omega_n) &=& -T\sum_{m,\vec{q_{\bot}}}
|\gamma(\vec{k_{\bot}}-\vec{q_{\bot}})|^2 D(\omega_m-\omega_n,
\vec{k_{\bot}}-\vec{q_{\bot}}) G(i\omega_m, \vec{q_{\bot}})
G(-i\omega_m, \vec{-q_{\bot}}) \Sigma_s(\vec{q_{\bot}}, i\omega_m),
\\
\Sigma^v_s(\vec{k_{\bot}}, i\omega_n)&=&
2T^2\sum_{m,l,\vec{q_{\bot}},\vec{p_{\bot}}}
|\gamma(\vec{k_{\bot}}-\vec{q_{\bot}})|^2
|\gamma(\vec{k_{\bot}}-\vec{p_{\bot}})|^2 D(\omega_m-\omega_n,
\vec{k_{\bot}}-\vec{q_{\bot}})D(\omega_n-\omega_l,
\vec{k_{\bot}}-\vec{q_{\bot}}) \nonumber \\
&\times& G(i\omega_l, \vec{p_{\bot}})
G(i(\omega_l-\omega_n+\omega_m),
\vec{p_{\bot}}-\vec{k_{\bot}}+\vec{q_{\bot}}) G(i\omega_m,
\vec{q_{\bot}}) G(-i\omega_m, \vec{-q_{\bot}})
\Sigma_s(\vec{q_{\bot}}, i\omega_m), \\
\Sigma^c_s(\vec{k_{\bot}}, i\omega_n)&=&
T^2\sum_{m,l,\vec{q_{\bot}},\vec{p_{\bot}}}
|\gamma(\vec{k_{\bot}}-\vec{p_{\bot}})|^2
|\gamma(\vec{q_{\bot}}-\vec{p_{\bot}})|^2 D(\omega_n-\omega_l,
\vec{k_{\bot}}-\vec{p_{\bot}})D(\omega_m-\omega_l,
\vec{q_{\bot}}-\vec{p_{\bot}}) \nonumber\\
&\times& G(i\omega_l, \vec{p_{\bot}})
G(i(\omega_l-\omega_n-\omega_m),
\vec{p_{\bot}}-\vec{k_{\bot}}-\vec{q_{\bot}}) G(i\omega_m,
\vec{q_{\bot}}) G(-i\omega_m, \vec{-q_{\bot}})
\Sigma_s(\vec{q_{\bot}}, i\omega_m), \nonumber\\
\end{eqnarray}
\end{widetext}
We take the average over Fermi energy($|\vec{k_{\bot}}| \approx
|\vec{q_{\bot}}| \approx k_f$), and denote
$\xi(\vec{p_{\bot}}-\vec{k_{\bot}}-\vec{q_{\bot}}) -\mu = x' +
2\epsilon_f\beta$, where
$$
\beta= 1+\cos(\phi')-\cos(\theta)-\cos(\gamma).
$$
We define the superfluid order parameter in the usual way,
$\Delta(i\omega_n)=\Sigma_s(i\omega_n)/Z(i\omega_n)$. Due to
single-component nature of the fermions in the mixture, the order
parameter can be expanded in odd partial wave
$\Delta(i\omega_n,\phi)= \sum_{L=...,-1,1,...}
\Delta_L(i\omega_n)\exp [ i L \phi ]$. In the strong coupling limit
we are interested in the terms $n=0$ and $n=-1$\cite{mahan}.
Assuming the order parameter to be even function of frequency,
$\Delta(\pi T)=\Delta(-\pi T)$, we get from Eq.~(\ref{s1}), (\ref{ss}),
\begin{equation}\label{su1}
\Delta_L(\pi T) Z(\pi T)= \lambda^{\Delta}_L(0,T)\Delta_L(\pi T)+\lambda^{\Delta}_L(-1,T)\Delta_L(-\pi T),
\end{equation}
where
\begin{eqnarray}\label{su2}
\lambda^{\Delta}_L(m,T)&=&\lambda_L(m,T)-2P_v(m,T)-P_c(m,T), \nonumber\\
\end{eqnarray}
$\lambda_L(m,T)$ originate from $\Sigma^1_s(i\omega_n)$, whereas
$P_v(m,T)$ comes from vertex corrected self-energy
$\Sigma^v_s(i\omega_n)$. $P_c(m,T)$ results from the contribution
the cross-term $\Sigma^c_s(i\omega_n)$,
\begin{widetext}
\begin{eqnarray}\label{pc}
P_c(m,T)&=& 2T {\rm Im} \left [ \sum_{l,\phi',\theta}
|\gamma(\theta)|^2|\gamma(\theta-\phi')|^2D(l, \theta-\phi')D(m-l, \phi') \right . \nonumber\\
&\times& \left .  \left [ \tan^{-1}\left
[\frac{\mu}{(2l+1)\pi T} \right ]-\tan^{-1}\left [\frac{\mu}{i\beta+(2l-2m-1)\pi T}
\right ] \right ]
\frac{1}{\beta+2i(m+1)\pi T} \right ]. \nonumber \\
\end{eqnarray}
\end{widetext}
Eq.~(\ref{su1}) is similar to the Eliashberg equation \cite{mahan} in
the strong coupling limit with additional vertex-corrected
interaction strengths. Apart from the dependance on temperature, the
effective interaction strength in the angular momentum channel $L$,
$\lambda^{\Delta}_L(T)$, is also a function of dimensionality $\eta$
and the boson-fermion mass ratio $m_f/m_b$. As noticed in
Eq.~(\ref{pvm}), $P_v(0,T)$ is positive,
thus reducing the effective interaction strength in the Cooper-pair channel. But the correction arising from
the cross term $P_c(m,T)$ can be negative or positive, as shown in Fig.~\ref{pc53}.  Also the sign of
%--------------------------------
%---------------------------------
$P_c(m,T)$ depends on the particular angular momentum channel under
consideration. But in general, for smaller $\eta$, $P_c(m,T)$
becomes positive, in turn reducing the interaction strength,
$\lambda^{\Delta}_L(m,T)$, in the Cooper-pair channel. But higher
value of $\eta$ changes $P_c(m,T)$ to negative, which enhances
$\lambda^{\Delta}_L(m,T)$. We also find out that with increasing
fermionic mass, $P_c(m,T)$ becomes negative for lower values of
$\eta$, thus enhancing the interaction strengths in the various
angular momentum channel. The reason behind this is that $P_c(m,T)$ depends on the ratio between
the position of the roton minimum $k_0$ and Fermi momentum $k_0/2k_f$.
With high fermionic mass, $k_f$ increases, resulting in lower ratio $k_0/2k_f$.
This in turn makes the cross term more negative. From this we infer that it is better to use fermions
with higher mass to get a higher interaction strength $\lambda^{\Delta}_L(m,T)$.

\begin{figure}[ht]
\begin{center} \epsfig{file=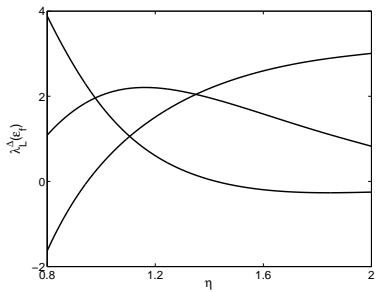,width=5cm} \caption{\label{etal}
Figure of effective interaction $\lambda^{\Delta}_L(\pi T=\epsilon_f)$ as a function of $\eta$ for $m_f/m_b=53/52$, $\pi T=\epsilon_f$, $g_{\rm 3d}=4.0$, $\mathcal{G}_{\rm bf}=.5$ and $\frac{3g}{4\pi g_{\rm
dd}}=.6$. The black, red and blue lines correspond to $\lambda^{\Delta}_1(T)$, $\lambda^{\Delta}_3(T)$, $\lambda^{\Delta}_5(T)$ respectively. }
\end{center}
\end{figure}

      Next we look into the dependance of $\lambda^{\Delta}_L(0, T)$ on dimensionality $\eta$. First
we fix $\pi T=\epsilon_f$ and plot for various value of $\eta$ n Fig.~\ref{etal}.
%--------------------------------
%---------------------------------
We see that the magnitude of $\lambda^{\Delta}_L(\epsilon_f)$ in
different angular momentum channel changes as $\eta$ varies. Also
different angular momentum channel becomes dominant depending on the
dimensionality. This qualitatively resembles the situation in
section 3, where without the vertex correction, we find that
different interactions becomes dominant for different
dimensionality.

      Next we find the transition temperature for fermionic superfluidity within perturbative scheme
as long as $T_c<T^*$ where $T^*$ is the temperature for which the
perturbative scheme becomes invalid. In the situation of more than one solution of Eq.~(\ref{su1}),
we have taken the transition temperature to be the one corresponding to the highest temperature. 
In the case of Eq.~(\ref{su1}) having no solution for $T_c$, vertex-corrected strong coupling
superfluidity is not possible and to find the transition temperature
we need to consider the full Eliashberg equation \cite{mahan} for
all values of $n$. This regime is not considered in this paper as we
are only interested in the high temperature limit of the transition
temperature.
\begin{figure}[ht]
\begin{center}
$\begin{array}{c} \\
\epsfig{file=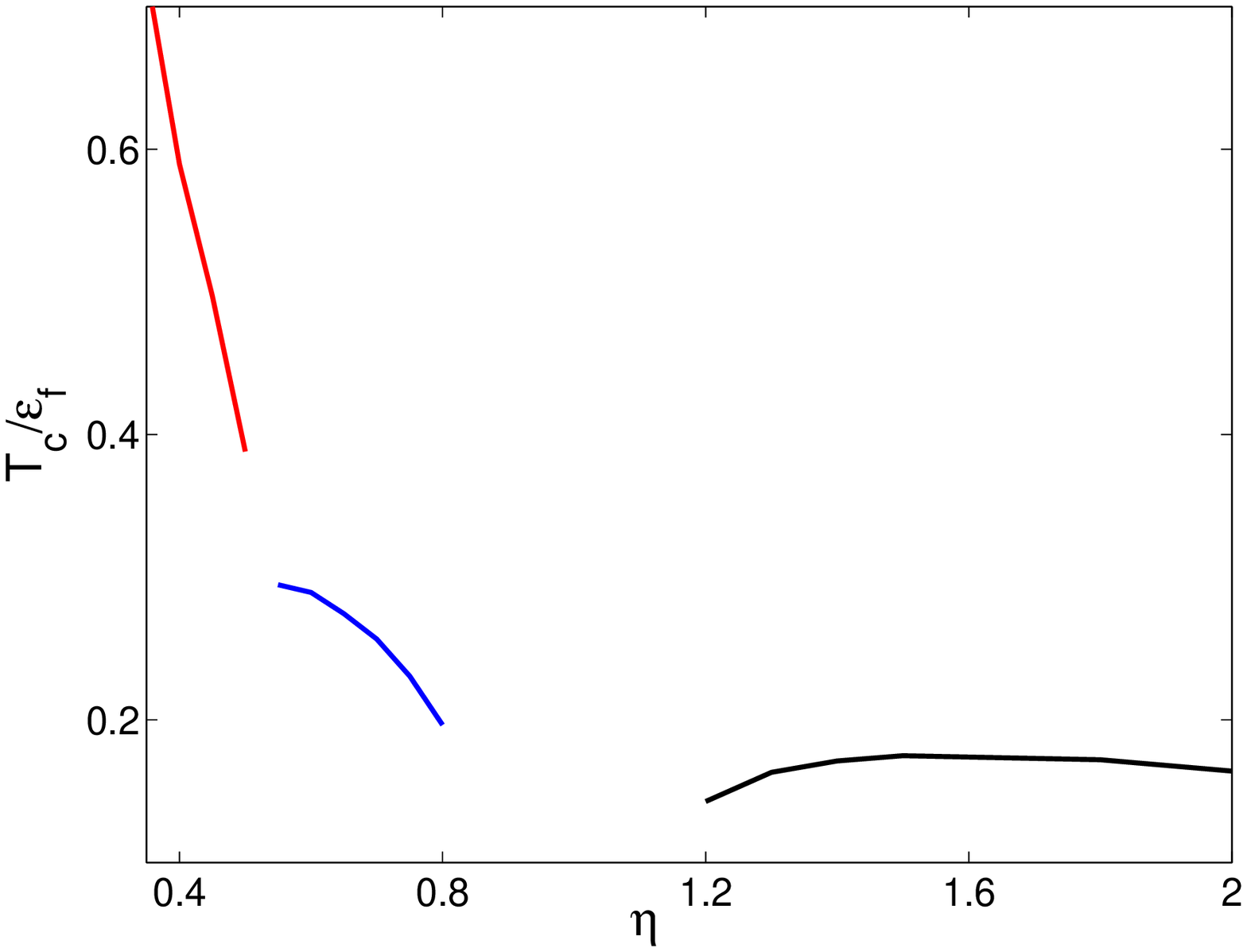, width=5.5cm} \\
{\mbox{\bf (a)}}\\
    \epsfig{file=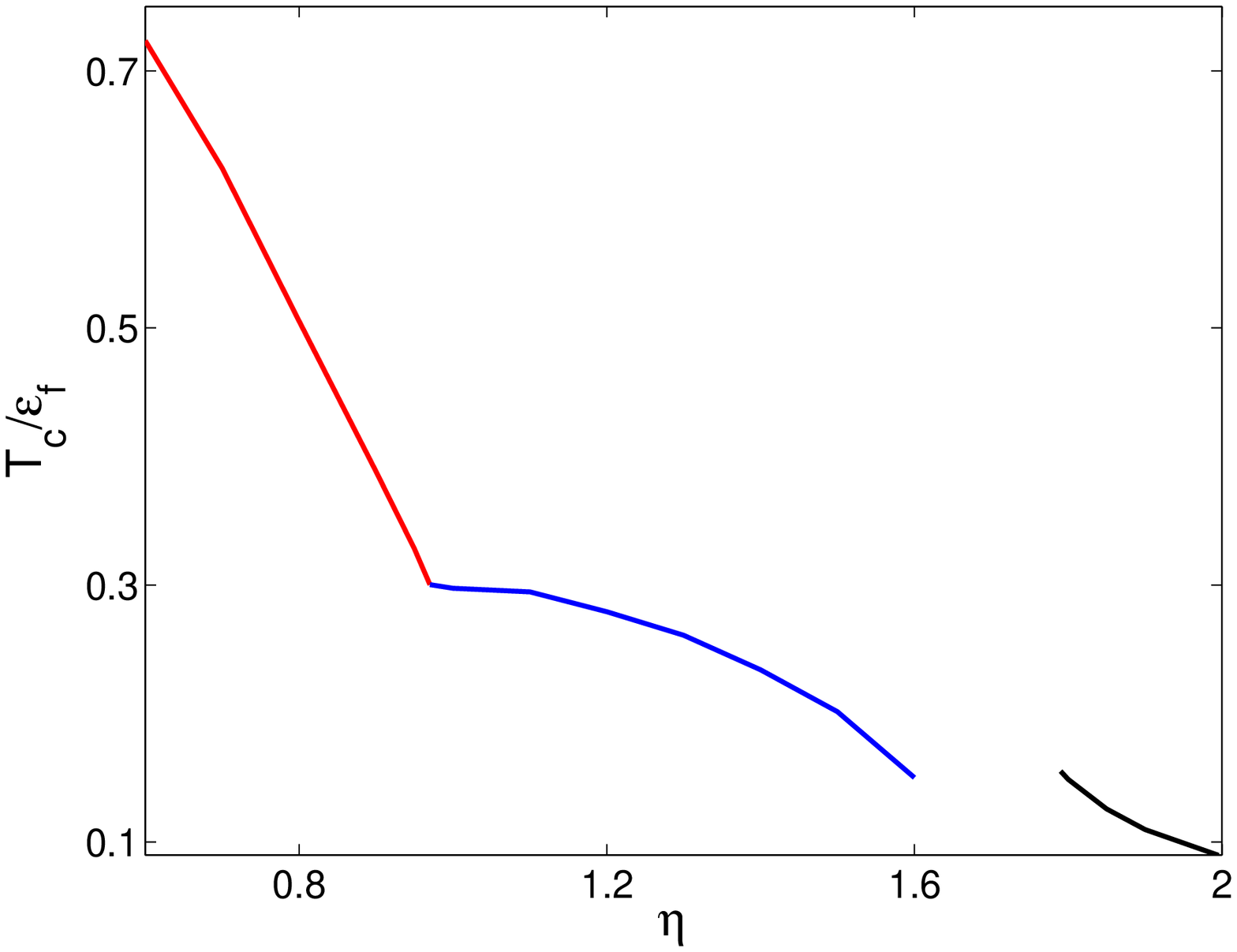, width=5.5cm} \\
    {\mbox{\bf (b)}} \\
\end{array}$
\caption{\label{templ}Critical temperature $T_c$, in the units of
Fermi energy $\epsilon_f$, as a function of $\eta$. The black, red
and blue lines correspond to transition temperature of $p$-wave,
$h$-wave, and $f$-wave order parameters respectively with parameters . a) For
parameters $m_f/m_b=1.7$, and dipole strength $g_{\rm 3d}=2.5$. b) For parameters
$m_f/m_b=53/52$, and dipole strength $g_{\rm 3d}=4.0$.}
\end{center}
\end{figure}
In Fig.~\ref{templ} (a) and (b) we plot the solution of the
Eqs.~(\ref{z11}), (\ref{su1}). In the case of obtaining multiple
solutions for transition temperature corresponding to different
angular momentum channel, we considered the channel with maximum
transition temperature to be the solution. In general we find that
with lowering dimensionality, the nature of the superfluidity
changes from $p$, to $h$ to $f$-wave. Also we obtained transition
temperature $T_c$ in the order of $.1\epsilon_f$ or more in each
angular momentum channel. By comparing Fig.~\ref{templ} (a) and (b)
we notice that, for higher fermion mass, we find that much lower
value of $\eta$ can be attainable without the perturbation becoming
invalid than for a lower fermionic mass.
%------------------------------------------------------------
\section{Chirality and non-Abelian Anyons}\label{abex}
In this section we briefly discuss the properties of quasi-particle excitations
inside a vortex for different internal symmetries of the order parameter.
By solving Bugoliubov-DeGennes equation in the limit of large distance from the 
core of the vortex, we show that the zero-energy solutions in the chiral $p-$, $f-$, $h-$ wave states are bounded
and non-Abelian in nature. 
   By considering the superfluid gap equation at low temperature, the
gap is maximum when the order parameter breaks time reversal
symmetry. From now on we assume that the order parameters are
denoted by $ \Delta_L=\Delta_0(\vec{r}) \left [ \frac{k}{k_f} \right
]^L e^{i L \theta}$ where $k_x=k\cos \theta$, $k_y=k\sin \theta$ and
$\Delta_0(\vec{r})$ is the center of mass amplitude of the Cooper
pairs, with $\vec{r}$ being the center of mass coordinate of the
pair. For vortex state $\Delta_0(\vec{r})$ cab be approximated as:
i) $\Delta_0(\vec{r})=0, r<\xi$ and ii) $\Delta_0(\vec{r})= \Delta_0
\exp(i\phi), r\geq\xi$, where $r=\sqrt{x^2+y^2}$, $\tan \phi=y/x$.
$\xi$ is the size of the core of the vortex. The vortex state of the
$p$-wave superfluids always has a zero-energy bound quasi-particle
state \cite{vol1, CN1, gur2}. Now we discuss the asymptotic
solutions for the zero-energy bound state for $f$- and $h$-wave
order parameters. The quasi-particle states in a single vortex can
be found in the limit of large distance from the vortex core by
solving the Bugoliubov-DeGennes equation,
%\begin{widetext}
\begin{eqnarray}\label{bdg}
H_0 u_L + (-i)^L \frac{\Delta_0}{k^L_f} e^{i\phi/2} \left [
e^{-i\phi} \left ( \frac{
\partial }{\partial r} - \frac{i}{r}\frac{ \partial
}{\partial \phi} \right ) \right ]^L e^{i\phi/2} v_L && \nonumber\\
=E u_L && \nonumber\\
- H_0 v_L  + (i)^L \frac{\Delta_0}{k^L_f} e^{-i\phi/2} \left [
e^{i\phi} \left ( \frac{
\partial }{\partial r} + \frac{i}{r}\frac{ \partial
}{\partial \phi} \right ) \right ]^L e^{-i\phi/2} u_L && \nonumber\\
= E v_L, && \nonumber\\
\end{eqnarray}
%\end{widetext}
where $E$ is the energy of the quasi-particles with amplitudes
$u_L,v_L$. We particularly look for zero energy solutions with
bounded $u_L,v_L$ with the property $u_L=v^{*}_L$ \cite{gur2}.
For $r\rightarrow \infty$, we can neglect the terms scaled as $r^{-1}$ in Eq.~(\ref{bdg}). Then the solution
of Eq.~(\ref{bdg}) with different orbital symmetries read,
\begin{equation} \label{angtop}
\bf \left[
\begin{array}{c}
u_1  \\
           \\
u_3  \\
                 \\
u_5               \\
\end{array}
\right] \sim \bf \left[
\begin{array}{c}
\exp \left ( -\frac{m_f \Delta_0}{k_f} r \right ) \\
\exp \left ( -\frac{k_f^3}{6 m_f \Delta_0} r \right ) e^{2i\phi} \\
\exp \left ( - \left [ \frac{k_f^5}{2 m_f \Delta_0} \right ]^{1/3}
r \right ) e^{4i\phi} \\
\end{array}
\right].
\end{equation}
The zero-energy solution for each odd-wave parameter corresponds to
different angular momentum channel of the quasi-particles inside a
vortex core. These results can also be carried out by applying the
``index theorem" \cite{tew}. For temperature smaller than the energy
gap $\Delta^2_0/\epsilon_f$, only the zero energy mode is occupied. The
quasi-particle operator in that situation is written as
$\gamma_L=\int d^2r (u_L(r)c^{\dagger}(r) + v_L(r)c(r))$, which acts
as Majorana fermion \cite{iva, CN1, gur2}.  $\gamma_L$ obeys
non-Abelian statistics and can be used for quantum computing
\cite{CN2}. In order to perform quantum computational tasks,
existence of several well separated vortices is necessary. We can
assume in the weak coupling limit, $\Delta_0=\beta \epsilon_f$, where
$\beta$ is a constant usually less than one. Then substituting
$\Delta_0$ in Eq.~(\ref{angtop}), we get $u_1 \propto \exp(-\beta
k_fr/2), u_3 \propto \exp(-k_fr/3\beta), u_5 \propto
\exp(-k_fr/\beta^{1/3})$. For $r\gg\xi$, for smaller $\beta$, $u_5$
has smaller tail than $u_3$ and $u_1$. Thus non-overlapped states
can be achieved more easily in superfluids in $L=5$ and $L=3$
channel than for p-wave channel.

\section{Conclusion}
 Summarizing, we studied boson-induced superfluidity of fermions in a
mixture of dipolar bosons and single-component fermions. A system is
proposed where conventional pairing mechanism gives rise to
different exotic internal structures of the Cooper pairs with strong
interaction in respective angular momentum channels. We find that
vertex-corrections play an important role for superfluidity in this
mixture and results in high value of transition temperatures. We
like to stress that the high transition temperatures  are a result
of the inclusion of vertex correction and cross interaction within
the Cooper-pair channel. Importantly, we find that by decreasing
$\eta$, we can generate exotic superfluids with $p$, $f$ and
$h$-wave internal structure. Excitations in these types of
superfluids breaks time-reversal symmetry and supports
quasi-particles with non-Abelian statistics.

\section{Acknowledgement}
 This work is financially supported by the Spanish MEC QOIT
(Consolider Ingenio 2010) projects, TOQATA (FIS2008-00784), MEC/ESF
project FERMIX (FIS2007-29996-E), EU IP grant AQUTE, ERC advanced grant QUAGATUA, and
Humboldt Foundation.


\begin{thebibliography}{99}
\bibitem{rz}
We determine the Thomas-Fermi radius $R_z$ by minimizing the
mean-field energy(with respect to $R_z$) of the bosons interacting
via $V_{\rm eff}(\tilde{k}_{\bot})$ .
\end{thebibliography}
\end{document}